\documentclass[twocolumn,showpacs,amsmath,amssymb,floatfix,aps]{revtex4}
\usepackage{graphicx}

\newcommand{\R}{\mathbb{R}}

\begin{document}

 \title{Early Universe Cosmology in Internal Relativity}
\date{\today}
    \author{Olaf Dreyer}
    \email{odreyer@mit.edu}
    \affiliation{Center for Theoretical Physics, Massachusetts Institute of Technology, 
    \\77 Massachusetts Ave, Cambridge, MA 02139}

\begin{abstract} We present a new approach to early universe cosmology. Inflation is replaced by a phase transition in which both matter and geometry are created simultaneously. We calculate the spectrum of metric perturbations and show that it is flat. We then argue that as a consequence of the dynamic nature of the phase transition the spectrum is likely not completely flat but tilted. We argue that the tilt is related to $\eta$, one of the critical exponents characterizing the phase transition. This exponent generically lies between $0.03$ and $0.06$. It thus coincides with the observed tilt of the perturbation spectrum. Because the critical exponent is related to the presence of an additional small length scale we argue that the deviation of the observed spectrum from flatness might be an experimental indication that our world is in fact discrete. 
\end{abstract}


\maketitle
\section{Introduction}\label{sec:intro}
With the advent of precision measurements of the microwave background cosmology has entered a new era. Cosmological theories are now very well constrained by experimental data. The current paradigm for early universe cosmology is Inflation. Inflation posits that in the early universe there was a period of exponential growth. Inflation predicts that as a consequence of this exponential grows the spectrum of metric perturbations $\Phi$ is given by
\begin{equation}\label{eqn:spectrum}
\delta^2_\Phi \sim k^{n_s-1},
\end{equation}
and $n_s$ is very close to unity. The spectrum of metric perturbations is thus almost flat. This spectrum is in fact seen by recent observations of the cosmic microwave background \cite{wmap}. The most likely range for the exponent is given by 
\begin{equation}\label{eqn:index}
0.02 \ <\  1-n_s\ < \ 0.05.
\end{equation}

\begin{figure}
\begin{center}
\includegraphics[height=4cm]{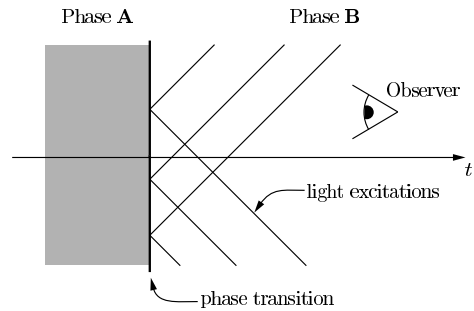}
\end{center}
\caption{In Internal Relativity a curved Lorentzian manifold is constructed from an underlying condensed matter model. The process that we are interested in in this paper is the emergence of the ordered phase \textbf{B} that looks like a Lorentzian manifold through a phase transition from an unordered phase \textbf{A}. In this article we investigate the consequences of the phase transition. How does the phase transition look to an observer in the ordered phase \textbf{B}?}
\label{fig:transition}
\end{figure}

In this paper we suggest a completely new paradigm for explaining this experimental data. This approach is based on a new theory of quantum gravity that we have called Internal Relativity \cite{dreyer1, dreyer2}. The starting point in Internal Relativity is a  condensed matter model like the Heisenberg model in three dimensions. Internal Relativity then asks and answers the question of how the physics of this model looks like when viewed from the point of view of an internal observer. It is because of this internal perspective that special relativity appears in a model that has a preferred time and a lattice as its foundation. In \cite{dreyer2} we have furthermore shown that generically an internal observer finds a curved Lorentzian manifold. In this paper we want to exploit the fact that the underlying model is a condensed matter model that can have different phases that are separated by a phase transition. Let us assume that the phase that looks like a curved manifold to an internal observer is characterized by an order parameter $\theta$. If this ordered phase came to exist through a phase transition from an unordered phase then we can ask how this transition looks like to the observer in the ordered phase (see figure \ref{fig:transition}). We claim that what this local observer will see when she looks towards the phase transition is in fact a flat spectrum.

The argument leading to this insight is rather straight forward. The main input from our quantum theory of gravity is the direct connection it gives between the order parameter $\theta$ and the Newtonian potential $\Phi$. In the cosmological context the Newtonian potential equals the scalar perturbation of the metric tensor. It is an advantage of Internal Relativity that it allows for this direct connection between a parameter in the fundamental theory and an observable quantity in the emergent theory. Given this connection we then just have to look at the behavior of the order parameter in the vicinity of a phase transition. This is a well studied question in condensed matter physics and we show that we arrive at the desired spectrum. 

The organization of the paper is now as follows. In the next four sections we review a number of topics that we touch on this paper. We start with a brief look at spectral analysis in section \ref{sec:spectral}. We pay particular attention to the connection between the spectral density and the two-point correlation function. Next we give a brief review of the basics of inflation in section \ref{sec:inflation}. We point out the basic mechanism with which inflation produces the spectrum (\ref{eqn:spectrum}) and take special note of the number of free parameters in the theory. We also have a look at what inflation has to say about the production of gravity waves. In section \ref{sec:ir} we review those parts of Internal Relativity that we need in this paper. In particular we focus on the relation between the order parameter $\theta$ and the Newtonian potential. Then we gather a few facts about second order phase transitions in section \ref{sec:pt}. Of special interest here is the behavior of the two-point correlation function near a phase transition. Finally we put all the ingredients together in section \ref{sec:cosmo}. Using the connection between the order parameter and the Newtonian potential we show that the spectrum of metric perturbations is exactly the one seen in the cosmic microwave background, i.e., the spectrum is the one given by equation (\ref{eqn:spectrum}). 

\section{Spectral analysis}\label{sec:spectral}
To calculate the spectral density $\delta_\theta^2$ of the order parameter we use one result from statistics. Let $f$ be a random field in $\R^3$. Then we can define two functions describing the statistic properties of $f$. The first one is the spectral density $h_f(k)$. It is given by
\begin{equation}
h_f(k) = \lim_{V\rightarrow\infty} \frac{\vert f_V(k)\vert^2}{V}.
\end{equation}
Here $f_V(x)$ denotes the function that is equal to $f$ if $x$ is in $V$ and zero otherwise. If $\chi_V$ is the characteristic function of $V$, then $f_V = \chi_V \cdot f$. The function $f_V(k)$ is the Fourier transform of $f_V$. 

The second function describing the statistic properties of $f$ is the is the two-point correlation function $G_f(x)$:
\begin{equation}
G_f(x - y) = \langle f(x) f(y) \rangle.
\end{equation}
We will restrict our attention to the homogeneous case in which $G$ is only depends on the differance between $x$ and $y$. It turns out that the correlation function $G_f(x)$ and the spectral density $h_f(k)$ are related by a Fourier transform.
\begin{equation}
h_f(k)  = G(k).
\end{equation}

The function that we will be most concerned with is the \emph{spectrum} or \emph{variance} $\delta_f^2$:
\begin{equation}
\delta_f^2(k) = \sqrt{\frac{2}{\pi}}\, k^3\, h_f(k).
\end{equation} 

In what follows we will often deal with functions that are powers of their arguments. In fact we will only be interested in the scaling exponent. Because of this it is convenient to note that if a function scales like 
\begin{equation}
f(x) \sim x^\alpha,
\end{equation}
then its Fourier transform will scale like 
\begin{equation}\label{eqn:fourierscaling}
f(k) \sim k^{-3-\alpha}.
\end{equation}

\section{Inflation}\label{sec:inflation}
According to inflation the universe underwent a period of exponential expansion during its early history. Inflation achieves this expansion by adding a new scalar field $\phi$ to the theory. This scalar field is called the inflaton. By introducing
\begin{eqnarray}
\varepsilon & = & \frac{1}{2}\partial^{\gamma}\phi\, \partial_{\gamma}\phi + V(\phi),\\
 & & \nonumber \\
 p & = & \frac{1}{2}\partial^{\gamma}\phi\, \partial_{\gamma}\phi - V(\phi),\\
 & & \nonumber \\
 u^\alpha & = & \partial^{\alpha} \phi\, (\partial^{\gamma}\phi\, \partial_{\gamma}\phi)^{-1/2},
\end{eqnarray}
where $V$ is the potential for $\phi$, the energy momentum tensor for the inflaton can be written like the energy-momentum for a perfect fluid:
\begin{equation}
T^{\alpha}{}_{\beta} = (\varepsilon + p) u^\alpha u_\beta - p\, \delta^{\alpha}{}_{\beta}
\end{equation}
In a homogeneous universe the spatial derivatives can be neglected. If furthermore the evolution of $\phi$ is slow so that the time derivative $\dot\phi$ can be neglected we have
\begin{equation}
\varepsilon = - p = V(\phi), 
\end{equation}
i.e. the inflaton acts like a cosmological constant. This is how the addition of $\phi$ leads to the exponential expansion. To achieve a smooth exit from this inflationary state the potential $V$ is chosen in such a way that it is not entirely flat but instead has a small slope that the inflaton rolls down. It turns out that only two parameters are required to capture the essential parts of the potential:
\begin{eqnarray}
\epsilon & = & \frac{1}{2}\left(\frac{V^\prime}{V}\right)^2\\
 & & \nonumber \\
 \eta & = & \frac{V^{\prime\prime}}{V}
\end{eqnarray}
These two parameters are the so called slow-roll parameters. They are free parameters of the theory; chosen to fit observation. 

\begin{figure}
\begin{center}
\includegraphics[height=4cm]{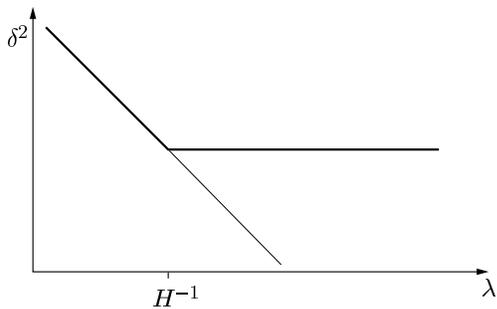}
\end{center}
\caption{Inflation predicts an almost flat spectrum of metric perturbations by introducing a new dynamical scale into the problem. This scale is the Hubble scale $H^{-1}$. The evolution of a mode depends on whether the wavelength $\lambda$ of the mode is bigger or smaller then the Hubble horizon. For $\lambda < H^{-1}$ the curvature of spacetime is of no importance. The spectrum of the mode is thus that of a free field in Minkowski space. For $\lambda>H^{-1}$ the evolution of the mode freezes and the amplitude stays essentially constant. The spectrum predicted by inflation is thus nearly flat. The fact that the exponential grows has to end implies that $H$ changes slightly and leads to a slight tilt of the spectrum. }
\label{fig:inflation}
\end{figure}

Inflation arrives at the spectrum in equation (\ref{eqn:spectrum}) by introducing a new dynamical scale into the problem. This new scale is the curvature radius during inflation, also called the Hubble scale, $H^{-1}$. The evolution of a mode in this expanding background depends crucially on whether the wavelength of the mode is smaller or larger then the Hubble scale. For a wavelength much smaller then the Hubble scale the curvature of spacetime is not important. The mode propagates as it would in Minkowski space. For wavelengths larger than the Hubble scale the curvature of spacetime becomes important. In fact the evolution of the mode effectively stops on these scales. During Inflation the physical wavelength of a given mode expands. Even if it starts out at scales smaller then the Hubble scale at some point it will cross that scale and effectively stop evolving. The amplitude of the mode after its wavelength has crossed the horizon is thus given by its amplitude when it crossed the horizon (see figure \ref{fig:inflation}).

If inflation were eternal the above mechanism would lead to a perfectly flat spectrum. Instead the small tilt of the potential $V$ leads to a small deviation of the spectrum from flatness. It is convenient to introduce the spectral index $n_s$ in terms of the spectral density by
\begin{equation}
n_s - 1 = \frac{d \ln \delta^2_\Phi}{d \ln k}.
\end{equation}
In terms of the slow-roll parameters $\epsilon$ and $\eta$ the spectral index for the simplest models of inflation is
\begin{equation}
n_s -1 = \frac{1}{4\pi}(\eta - 3\epsilon).
\end{equation}
(see \cite{slava} for more details.) Equation (\ref{eqn:index}) gives the typical range for the spectral index when using the above formula.

Another free parameter of inflation is the overall amplitude of the observed spectrum. The microwave background is isotropic to one part in $10^5$. Inflation does not provide an argument for this number. It too, together with the slow-roll parameters, has to be fixed from the outside. 

So far we have concentrated on the scalar perturbations induced by the presence of the inflaton. The same arguments that give the spectrum of these scalar perturbations can be applied to calculate the spectrum of the perturbations of the metric itself. These tensor perturbations are weaker then the scalar perturbations and have thus far not been  observed. If we denote by $\delta^2_h$ the spectrum of the metric perturbations $h$ then one obtains for one scalar field and the simplest model of inflation (again, see \cite{slava} for details)
\begin{equation}\label{eqn:tensor}
\frac{\delta^2_\Phi}{\delta^2_h} = 0.2\, \ldots\, 0.3.
\end{equation}
The metric perturbations are thus roughly an order of magnitude weaker than the scalar perturbations. This is the reason why they have not yet been observed. Future experiments will be sensitive enough to see the metric perturbations. 

\section{Internal relativity}\label{sec:ir}
The most important feature of Internal Relativity is that the curved metric is obtained solely from the matter content of the emergent theory. In the context of a theory with an order parameter $\theta$ we have shown \cite{dreyer2} that the gravitational mass $m$ of a bound object is given by 
\begin{equation}
m \sim \int_{\partial  \mbox{\scriptsize \textbf{\textsf c}}} (\nabla\theta)\cdot d\sigma, 
\end{equation}
where $\partial\mbox{\textbf{\textsf c}}$ is a contour enveloping the object. This formula shows that the order parameter $\theta$ is proportional to the Newtonian potential. The metric in the presence of a non-zero order parameter is then
\begin{equation}
ds^2 = (1 + \kappa\theta) d\eta^2 - (1 - \kappa\theta)\delta_{ij} dx^i dx^j,
\end{equation}
where $\kappa$ is a proportionality constant that is of no concern to us right now. 
From this expression we see that we can identify $\theta$ with the function $\Phi$ that characterizes scalar metric perturbations in the longitudinal (conformal-Newtonian) gauge
\begin{equation}
\theta \sim \Phi.
\end{equation}
To find the the spectrum $\delta_\Phi^2$ of the metric perturbations we then just have to find the spectrum $\delta^2_\theta$ of the order parameter:
\begin{equation}
 \delta_\theta^2 \sim  \delta_\Phi^2
\end{equation}
From section \ref{sec:spectral} we know that we can find $\delta^2_\theta$ by looking at the two-point correlation function for $\theta$. 

Because in Internal Relativity the metric is obtained from the emergent matter there exists no metric field that can fluctuate on its own. Inflation produces measurable tensor perturbations of the metric by amplifying the zero mode fluctuations of the metric field. These independent metric fluctuations do not exist in Internal Relativity and it follows that the spectrum of tensor metric perturbation is far smaller than the ratio given in equation (\ref{eqn:tensor}). We note that in this respect Internal Relativity makes the same prediction as the computational universe \cite{seth} and quantum graphity \cite{graphity1, graphity2}.

\begin{table}
\caption{Behavior of the correlation function $G(x)$.}\label{table:correlation}
\begin{ruledtabular}
\begin{tabular}{cccc}
 &Temperature &G(x)&  \\
 \hline
 & & & \\
& $T\, >\, T_c$ &$ e^{-\vert x\vert/\xi}\,\vert x \vert^{-d+2-\eta}$& \\
  & & & \\
& $T\, =\, T_c$ &$\vert x \vert^{-d+2-\eta}$ & \\
 & & & \\
 \hline
\end{tabular}
\end{ruledtabular}
\end{table}

\section{Phase transitions}\label{sec:pt}
In this section we want to review the behavior of the two-point correlation function $G(x)$ for the order parameter $\theta$ in the vicinity of a second-order phase transition. We start in the disordered phase with a temperature $T$ that is larger than the critical temperature $T_c$. This phase is characterized by a correlation length $\xi$. Values of  $\theta$ that are separated by more than the correlation length are essentially uncorrelated. The correlation function has the general form
\begin{equation}\label{eqn:subcritical}
G_{T>T_c}(x) \sim \frac{e^{-\vert x\vert/\xi}}{\vert x \vert^{d-2}},\ \ T\, >\, T_c
\end{equation}
where $d$ is the spatial dimension of the system. In the examples we are interested in here we have $d=3$. As the temperature $T$ approaches the critical temperature $T_c$ the correlation length $\xi$ diverges. The divergence of $\xi$ is described by some critical exponent $\nu>0$ and is of the form
\begin{equation}
\xi \sim \left\vert \frac{T - T_c}{T_c}\right\vert^{-\nu}.
\end{equation}
When the temperature $T$ reaches the critical temperature $T_c$ the exponential in equation (\ref{eqn:subcritical}) becomes unity and the correlation function now reads
\begin{equation}\label{eqn:critical}
G_{T=T_c}(x) \sim \frac{1}{\vert x \vert^{d-2}},\ \ T\, =\, T_c.
\end{equation}
When the temperature falls further the system transitions to the ordered state. The correlation function is now simply given by a constant.
\begin{equation}\label{eqn:aftertrans}
G_{T<T_c}(x) = \text{const.} \sim \vert x\vert^0
\end{equation}
It turns out that the above behavior of the correlation function is somewhat too simple. During the phase transition fluctuations on all length scales are important. Even a tiny length scale like a lattice constant in a condensed matter system will show up by slightly changing the exponents in the above expressions for the correlation function. The actual behavior of the correlation function is given in table \ref{table:correlation}. All exponents are adjusted by a small deviation $\eta$. We have gathered the typical values for this critical exponent $\eta$ in table \ref{table:eta} (this table is an adaptation of a table in \cite{chaikin}). 

\begin{table}
\caption{The critical exponent $\eta$. }. \label{table:eta}
\begin{ruledtabular}
\begin{tabular}{clcc}
 & System & $\eta$ & \\
 \hline
 & & & \\
 & Mean-field & 0 & \\
 & & & \\
 & 3D theory & & \\
 & n=1 (Ising) & 0.04 & \\
 & n=2 ($xy$-model) & 0.04 & \\
 & n=3 (Heisenberg) & 0.04 & \\
 & & & \\
 & Experiment & & \\
 & 3D n=1 & 0.03 -- 0.06 \\
 & & & \\
 \hline
\end{tabular}
\end{ruledtabular}
This table has been adapted from \cite{chaikin}. $n$ is the dimension of the order parameter.
\end{table}

\section{Cosmology}\label{sec:cosmo}
In this section we now investigate what Internal Relativity has to say about early universe cosmology. The situation that we are imagining is the one discussed in the introduction and depicted in figure \ref{fig:transition}. We are looking at a system that undergoes a second order phase transition. During the phase transition the system goes from a disordered phase to an ordered phase. We are interested in the traces the phase transition leaves in the ordered phase. How does the phase transition look like for an observer in the ordered phase looking back at the transition? Before we answer this question we remark that this scenario immediately solves the horizon problem.

\begin{figure}
\begin{center}
\includegraphics[height=5cm]{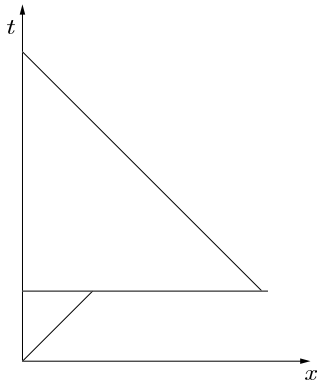}
\end{center}
\caption{The horizon problem. The microwave background is uniform to one part in $10^5$. This suggests that all parts of the surface of last scattering have been in causal contact with each other at some point. In standard pre-inflation cosmology this is not the case. This diagram shows that the particle horizon at recombination is much smaller then the part of surface of last scattering that is visible to us.}
\label{fig:horizon}
\end{figure}

The microwave radiation that we observe today traveled through the universe essentially unobstructed since it last scattered when the universe became transparent at the time of recombination. As mentioned above, the cosmic microwave background is uniform to about one part in $10^5$. This suggests that all parts of the surface of last scattering have been in causal contact at some point to achieve this uniformity. In pre-inflationary cosmology this was unfortunately not the case (see figure \ref{fig:horizon}). The great uniformity of the microwave background thus becomes a problem: the horizon problem. In our scenario this conundrum is immediately resolved. The horizon problem as we have stated it here requires for its formulation that the notion of locality as it is given by the local light cones never changes. In our scenario this is precisely not the case. The notion of locality that we infer in Internal Relativity through the light excitations of our model is only available in the ordered phase. There is no sense of locality in the unordered phase. Furthermore we have seen in section \ref{sec:pt} on phase transitions that the correlation length $\xi$ diverges at the phase transition. All parts of the system become correlated. The horizon problem thus seizes to be a problem in our setup. 

We now turn our attention to the spectrum. We want to calculate the spectrum $\delta^2_\Phi$ of the metric perturbations. In section \ref{sec:ir} on Internal Relativity we have seen that the metric perturbation $\Phi$ is proportional to the order parameter $\theta$. We thus have to calculate $\delta^2_\theta$. From section \ref{sec:spectral} we know that the spectral density $\delta^2_\theta$ is related to the two-point correlation function $G(x)$. We are interested here in the form of the correlation function \emph{after} the phase transition. From equation (\ref{eqn:aftertrans}) we know what the form of the correlation function is. Using also equation (\ref{eqn:fourierscaling}) we find:
\begin{eqnarray}
{\tilde\delta}^2_\Phi(k) & \sim & {\tilde\delta}^2_\theta(k)\\
 & \sim & k^3\, G_{T<T_c}( k ) \\
 & \sim & k^3 \frac{1}{k^3}\\
 & = & \text{const.}
\end{eqnarray}
We thus find a flat spectrum. In the next section we investigate what this picture has to say about the observed tilt of the spectrum. 

\section{The tilt}\label{sec:tilt}
In the preceding section we have found a flat spectrum of metric perturbations. Observations show that the spectrum is not exactly flat but is of the form given in equation (\ref{eqn:spectrum}). In our derivation of the spectrum we have used the correlation function $G_{T<T_c}$. When one compares it to $G_{T=T_c}$ and $G_{T>T_c}$ one sees that the critical exponent $\eta$ makes no appearance. This is no coincidence because $G_{T<T_c}$ is the correlation function for the ground state of the model. It retains no memory of the fact that the ground state was reached via a phase transition. Could it be that the correlation function $G_{T<T_c}$ from equation (\ref{eqn:aftertrans}) is too simple in that it does not accurately reflect the origin of the state in a phase transition? Comparing the range of the critical exponent $\eta$ from table \ref{table:eta} with the observed value of the spectral tilt in equation (\ref{eqn:index}) this is indeed an intriguing possibility.

The question that we are asking here is unfortunately a question that is concerned with the dynamic nature of the actual phase transition. We are asking what happens when we change the temperature of the system from $T=T_c$ to $T=T_c-\epsilon$ for some small $\epsilon$? Unfortunately, this type of question has received not nearly as much attention as the behavior of the system before and at the phase transition. The reason for this is that this last part of the transition is governed by the highly non-trivial dynamics of the full system and does not admit any simplifying assumptions. The answer to this question is also not likely to be universal. It is conceivable that the exact form of the correlation function after the phase transition depends on how the system underwent the transition. It might matter whether the transition was fast or slow.

In the absence of a better understanding of a phase transition we are forced to rely on arguments and conjectures. We first note that the correlation functions $G_{T=T_c}$ and $G_{T<T_c}$ are both scale free. Both are just powers of $k$. We would assume that whatever the modification of $G_{T<T_c}$ due to the transition is, it does not spoil this property. We thus expect that the real $G_{T<T_c}^\prime$ is also just a power of $k$:
\begin{equation}
G_{T<T_c}^\prime(k) \sim k^\gamma,
\end{equation}
for some exponent $\gamma$. Since we have $\gamma=0$ for the true ground state we can assume furthermore that $\gamma$ is small:
\begin{equation}
\vert\gamma\vert \ll 1
\end{equation}
In the next step we link $\gamma$ to the critical exponent $\eta$. The form of the correlation functions $G_{T>T_c}$ and $G_{T=T_c}$ in equations (\ref{eqn:subcritical}) and (\ref{eqn:critical}) follows from assumptions about the isotropy of the free energy of the system \cite{huang}. The critical exponent $\eta$ indicates a deviation of the actual behavior of the correlation function from this simple behavior that is due to the presence of another scale, namely the lattice spacing. Near the phase transition fluctuations of all scales occur and the presence of the lattice scale up in the critical exponent $\eta$. We now assume that the influence of the lattice is still felt in the final transition and that the effect is roughly of the same size as before the transition:
\begin{equation}
\vert\gamma\vert \simeq \eta.
\end{equation}
This leaves us with the question of the sign of $\gamma$. The sign of $\eta$ in table \ref{table:correlation} is negative which implies that $\eta$ will appear with a positive sign in the Fourier transform of the correlation function (see equation (\ref{eqn:fourierscaling})). We thus have to assume that the dynamics effect of the transition is such that it reverses the sign of the exponent. In this case we have
\begin{equation}
\gamma \simeq -\eta. 
\end{equation}
It is somewhat disappointing that the connection between $\gamma$ and $\eta$ can not be established more forcefully. The possibility of such a connection is worth mentioning though because it would be a strong indication that the observed spectral tilt of the microwave background radiation is a sign of a fundamental discreteness in nature.

It is not clear that the problem encountered above can be resolved using theoretical methods. One should be able though to clarify the situation using laboratory experiments. In this paper we are proposing to think of the universe as a solid state model. In the analysis above we use very generic features of such solid state models. The question of the behavior of the correlation function $G_{T<T_c}$ should be verifiable in a table-top solid state experiment.   

\section{discussion}\label{sec:discussion}
Inflation has been very successful in describing the observed spectrum of the microwave background. In this paper we are proposing to replace inflation with a phase transition. The nature of the phase transition is rather radical in that it creates the smooth spacetime that we find around us. Before this transition our familiar notions of space, time, and locality do not exist. Such a transition is possible in Internal Relativity, a new approach to the problem of quantum gravity. In Internal Relativity geometry is derived from the dynamic behavior of matter. If the matter is emergent then so is the geometry. Here we have shown that the effect of such a transition is a flat spectrum of metric perturbations:
\begin{equation}
{\tilde\delta}^2_\Phi(k) = \text{const.}
\end{equation}
We have furthermore argued that because of the dynamic nature of the phase transition the spectrum is actually of the form
\begin{equation}
{\tilde\delta}^2_\Phi(k) \sim k^{-\eta},
\end{equation}
where $\eta$ is a critical exponent. Typical values for this critical exponent are given in table \ref{table:eta}. With these values the spectrum calculated here becomes a very good fit with the observed spectrum of equations (\ref{eqn:spectrum}) and (\ref{eqn:index}). 

An intriguing aspect of the above picture is that it gives an indication of a fundamental discreteness of nature. The presence of the critical exponent $\eta$ is usually understood as indicating the presence of another length scale in the problem. In ordinary solid state models this is the lattice spacing. In our setting it indicates a new fundamental scale. The tilt of the microwave background spectrum can then be viewed as an observation of discreteness in nature. 

An uncertain point in the above argument is the exact behavior of the correlation function $G_{T<T_c}$. We have argued that the behavior of $G_{T<T_c}$ is such that we obtain the desired spectrum. Unfortunately we are not able to prove that this is the case. It is also not clear whether such a proof is possible. This is because the behavior that we are interested in is the non-universal behavior of a quantum many-body system. It seems possible though to investigate this problem using laboratory solid-state experiments. 

Because we are radically changing notions of locality we can also address the horizon problem in cosmology. Before the phase transition our current notion of locality does not exist because the matter degrees that define locality have not emerged yet. This change of locality can be compared to the change of locality in \cite{moffat} and \cite{aajm}. In \cite{moffat, aajm} the authors discuss the effects of a varying speed of light. Our approach is more radical in that we discuss the emergence of light. It is not that the properties of light are changing but that light does not exist in one phase of the system and only appears after the phase transition. 

The horizon problem is also tackled in quantum graphity \cite{graphity1, graphity2}. In this work space is created starting from a complete graph, i.e. a set of points that are all connected to each other. The authors then provide a Hamiltonian that erases edges from this complete graph to end up with a regular graph as a discrete model of a smooth spacetime. Since all points of space have been in contact at one point the horizon problem does not arise. It is not yet clear how different this point of view is to the one presented in this paper. One thing to note is that ultimately the geometry in quantum graphity is obtained internally through the available degrees of freedom. The system will have to go through a matter creating transition like the one described here after the basic lattice has been created. The arguments presented here should thus apply to this second transition and it is not clear whether the first transition to the regular lattice makes any discernible contribution. 

We also have to compare our approach to Inflation. Inflation has been extremely successful in describing the spectrum of the microwave background. It does so at the cost of introducing a hitherto unknown scalar field together with a potential that can be adjusted to fit the observations. As we have seen in section \ref{sec:inflation} the total amount of freedom can be reduced to three parameters: the two slow-role parameters together with the overall amplitude of the perturbations. Compared to this our scenario is rather more constrained. The numbers we use are those of generic solid state systems. The critical exponent $\eta$ is always in the narrow range seen in table \ref{table:eta} and can not be freely adjusted. The price to pay is of course a rather big change in the basic setup of the fundamental setup of the theory. 

The idea to replace inflation with a phase transition has appeared before in the context of string theory \cite{brandenburger}. Here the transition is from a Hagedorn phase to the radiation-dominated phase of standard cosmology. The extended nature of strings is responsible for the correct scaling of the resulting metric perturbations. The phase transition discussed here is quite different from the one we have in mind though. In \cite{brandenburger} cosmological perturbation theory is used throughout and the notion of a spacetime is always applicable. In our approach the notion of a spacetime geometry is only applicable after the phase transition. We argue that it is the creation of spacetime itself that gave rise to the metric perturbations rather then the particular behavior of certain matter in an already existing spacetime.

We finally want to comment on \cite{mag2}, another approach to early universe cosmology that makes use of a phase transition. The phase transition described in\cite{mag2} is a transition \emph{within} a quantum theory of metrics. The authors argue that such a quantum theory of metrics has at least two phases. One phase is well described by a classical metric that obeys the Einstein equations and the other phase is a high temperature phase that is a quantum superposition of metrics that can not be well described by just one classical metric. The claim is then that the observed spectrum of the microwave background radiation is a relic of the transition from the high to the low temperature phase. 

One difference to the mechanism presented in this paper is that the theory is based on a quantum theory of metrics. The Einstein action is assumed to be valid even in the high temperature phase. Internal relativity is different in that the presence of a metric is tied to the presence of matter. Fundamentally it is not a quantum theory of metrics. 

Another problem with a quantum theory of metrics, i.e. with a quantization of gravity, is that so far we have failed to construct such a theory. We know very little about how to construct the theory, investigate its phase diagram, or describe transitions between the different phases. In \cite{mag2} this problem is circumvented by appealing to the holographic principle in the high temperature phase. If we understood the theory we should be able to derive such a principle from it. Without this understanding the principle is very much disconnected from the fundamental setup.

\begin{acknowledgments} I would like to thank S.~Lloyd for inspiring discussions. I have to especially thank F.~Markopoulou. For a long time she has advocated the idea that notions of locality can change. To apply this idea to cosmology originated in a discussion with her. I would also like to thank R.~H.~Brandenberger, J.~Magueijo, J.~Pullin, and L.~Smolin for their comments on an earlier draft of the paper. Finally I would like to thank FQXi for funding this research and the W. M. Keck Foundation Center for Extreme Quantum Information Theory for hosting me at the Massachusetts Institute of Technology. 
\end{acknowledgments}


\begin{thebibliography}{99}
\bibitem{wmap} J.~Dunkley et al., \emph{Five-Year Wilkinson Microwave Anisotropy Probe (WMAP) Observations: Likelihoods and Parameters from the WMAP data}, e-Print: 0803.0586v1. 
\bibitem{dreyer1} O.~Dreyer, \emph{Emergent general relativity}, Contribution to book 'Towards Quantum Gravity'. Edited by D. Oriti. Cambridge University Press, 2006. e-Print: gr-qc/0604075
\bibitem{dreyer2} O.~Dreyer, \emph{Why things fall.} Proceedings of 'From Quantum to Emergent Gravity: Theory and Phenomenology', Trieste, Italy, 11-15 Jun 2007, PoS(QG-Ph)016, e-Print: arXiv:0710.4350.
\bibitem{slava} V.~Mukhanov, \emph{Physical foundations of cosmology}, Cambridge University Press, 2005. 
\bibitem{seth} S.~lloyd, \emph{The Computational universe: Quantum gravity from quantum computation.}, e-Print: quant-ph/0501135.
\bibitem{graphity1}T.~Konopka, F.~Markopoulou, L.~Smolin, \emph{Quantum graphity}. e-Print: hep-th/0611197.
\bibitem{graphity2}T.~Konopka, F.~Markopoulou, S.~Severini, \emph{Quantum Graphity: A Model of emergent locality.}. e-Print: arXiv:0801.0861.
\bibitem{chaikin} P.~M.~Chaikin and T.~C.~Lubensky, \emph{Principles of condensed matter physics}, Cambridge University Press, 2000.
\bibitem{huang} K.~Huang, \emph{Statistical mechanics}, John Wiley \& Sons, 1987.
\bibitem{moffat} J.~W.~Moffat, \emph{Superluminary Universe: A Possible Solution to the Initial Value Problem  in Cosmology}, Int. J. Mod. Phys. D\textbf{2}(3), 351 (1993).
\bibitem{aajm} A.~Albrecht and J.~Magueijo, \emph{Time varying speed of light as a solution to cosmological puzzles}, Phys. Rev. D\textbf{95}(4), 043516 (1999). 
\bibitem{brandenburger} A.~Nayeri, R.~H.~Brandenberger, C.~Vafa, \emph{Producing a Scale-Invariant Spectrum of Perturbations in a Hagedorn Phase of String Cosmology}, Phys. Rev. Lett. \textbf{97}(2), 021302 (2006).
\bibitem{mag2} J.~Magueijo, L.~Smolin, and C.~R.~Contaldi, \emph{Holography and the scale invariance of density fluctuations}, Class. Quantum Grav. \textbf{24}(14), 3691 (2007). 

\end{thebibliography}
\end{document}